\documentclass[prd,showpacs,amsmath,showkeys,twocolumn,floatfix,amssymb, preprintnumbers, nofootinbib, superscriptaddress]{revtex4} 
\usepackage{epsfig,dcolumn}
\usepackage{graphicx}
\usepackage{comment} 
\usepackage{verbatim}
\usepackage[usenames]{color}
\usepackage{graphicx}
\usepackage{bm}
 \usepackage{ifpdf}

\def\B{{\bf B}}

\def\lsim{\mathrel{\rlap{\lower4pt\hbox{\hskip1pt$\sim$}}
    \raise1pt\hbox{$<$}}}
\def\gsim{\mathrel{\rlap{\lower4pt\hbox{\hskip1pt$\sim$}}
    \raise1pt\hbox{$>$}}}
\begin{document}

\title{A coupled-channel formalism for   three-body final state interaction    }

\author{Peng~Guo}
\email{pguo@jlab.org}
\affiliation{Physics Department, Indiana University, Bloomington, IN 47405, USA}
\affiliation{Center For Exploration  of Energy and Matter, Indiana University, Bloomington, IN 47408, USA.}
\affiliation{Department of Physics and Engineering,  California State University, Bakersfield, CA 93311, USA.}

\date{\today}

\begin{abstract} 
  From   dispersion relation approach, a formalism that describes    final state interaction among three particles   in a coupled-channel system is presented.  Different representations of coupled-channel three-body formalism  for spinless particles in both initial and final states  are derived.
 
  \end{abstract} 

\pacs{ }

\maketitle


{\it Introduction.}---Hadron spectroscopy is one of   important methods for studying non-perturbative QCD  and gaining insights of hadron structures and  decay mechanism. With  high statistic data collected from facilities, such as BESIII, Jefferson Lab and Panda, data analysis becomes even more challenging than ever before, especially, multiparticle dynamics may play the central pole in inelastic region, without proper consideration of   constraints from   physics principles and multiparticle dynamics,   amplitudes extracted from data may be misleading. Therefore, to understand   phenomena precisely,   theoretical description of  decay  amplitudes need to take   into account all   possible dynamics, and   follow  some basic physics principles, such as unitarity and analyticity.  In the past,   handling processes with multiple-particle final states  has been mainly based on the  isobar model \cite{Goradia:1975ec,Ascoli:1975mn}, {\it i.e.} assuming a multiparticle decay  proceeded through a series of  quasi-two-body sequential decays. For  example, a decay process  of   one particle (0)   into three final states (1,2,3)      is usually described by a sum of all possible decay chains: $0 \rightarrow (12)3+1(23)+(31)2 \rightarrow 123$. For each individual decay chain, the amplitude is a   product of  kinematic factors, a coupling constant and a two-body     amplitude that only depends on  two-particle subenergy.
 Interaction among multiple final state particles has been ignored completely in the isobar model.

Three-body correction to isobar model has been developed  since 1960 \cite{Khuri:1960zz,Bronzan:1963kt,Aitchison:1965kt,Aitchison:1965dt,Aitchison:1966kt,Pasquier:1968zz,Pasquier:1969dt,Dodd:1977pj,Kambor:1995yc,Schneider:2011dt,Schneider:2012ez,Guo:2014vya,Guo:2014mpp,Danilkin:2014cra,Guo:2015zqa}, which is based on subenergy dispersion relation approach by considering the unitarity and analyticity properties of amplitudes. In   those dispersive approaches  \cite{Aitchison:1965kt,Aitchison:1965dt,Aitchison:1966kt,Pasquier:1968zz,Pasquier:1969dt,Kambor:1995yc,Schneider:2011dt,Schneider:2012ez,Guo:2014vya,Guo:2014mpp,Danilkin:2014cra,Guo:2015zqa},  a decay amplitude  is  written   as the sum of all possible decay chains.  For  each individual decay chain, the amplitude now is the  product of  kinematic factors, a subenergy dependent complex scalar function.
This scalar function satisfies  a coupled dispersion relation equations, and the solutions of these equations describe the rescattering effects among three particles. In this approach,  interaction among three particles is generated from pair-wise two-body interactions by exchanging a particle between   pairs. The unitarity and analyticity  are  guaranteed naturally.  
However, all the previous developments have not considered the contribution from inelastic channels  yet. In reality, the subenergy in the most of hadron production  processes usually is far beyond the elastic region. Once inelastic channels open up,  the interference between different channels may be important \cite{Guo:2010gx,Guo:2011aa}.  In recent years, the demand for studying and including three-body effect has been increased significantly, such as, for excited baryon study at Jefferson Lab. The complication for  establishing higher excited baryon states in those studies are not only  because most of those baryon states are produced from  multiple-particle final states but also  from the strongly coupled multiple channels in inelastic region.  Similar situation may exist in  incoming  exotic mesons studies at Hall D, Jefferson Lab and ongoing excited charmonium studies at BES III. Therefore, to disentangle all the coupled-channel effects from the multiple-particle final state interaction, a coupled-channel formalism for multiple-particle states is essential, some efforts based on effective theory formalism have been made along this line \cite{Kamano:2011ih,Nakamura:2012xx,Nakamura:2015qga}. The goal of this work is to   generalize dispersive   three-body rescattering formalism   to include the channels in inelastic region. In this letter, the decay process of a spinless-particle to three spinless-particle is presented to demonstrate the basics of coupled-channel three-body formalism without complication of spin structure of particles.

{\it Basic representation of coupled-channel three-body formalism.}---The  decay   of  a spinless particle    to three spinless particles  is described by,
\begin{equation}
\langle   1(\alpha) 2(\beta) 3 (\gamma) , \mbox{out}|  0, \mbox{in} \rangle= i (2\pi)^{4} \delta^{4} (\sum_{i=1,2,3} p_{i}-P) T^{\alpha \beta \gamma} ,
\end{equation}
where $(\alpha,\beta,\gamma)$ stand for the species of final state particles $(1,2,3)$     respectively,  and     the four momenta of i-th final state particle and the parent particle are denoted by   $p_{i}$ and $P$ respectively.  The   decay amplitude $T^{\alpha \beta \gamma} $  are usually expressed as the sum of partial wave series in each  two-body subenergy-channel \cite{Khuri:1960zz,Bronzan:1963kt,Aitchison:1965kt,Aitchison:1965dt,Aitchison:1966kt,Pasquier:1968zz,Pasquier:1969dt},
\begin{align}\label{decayamp}
& T^{\alpha \beta \gamma} =       \sum_{ L}     (2L+1)  \left [ P_{L}(z_{\alpha \beta}) F^{(\alpha \beta)}_{ L} (s_{12})   \right. \nonumber \\
&  \quad \quad  \quad  \left. + P_{L}(z_{ \beta \gamma}) F^{( \beta \gamma)}_{ L} (s_{23}) + P_{L}(z_{\gamma \alpha  }) F^{(\gamma \alpha  )}_{ L} (s_{31}) \right ],
\end{align}
where the isospin couplings have been suppressed for simplification purpose    only,     the invariants are defined by \mbox{$s_{ij} =(p_{i}+ p_{j})^{2}$} and three invariants are constrained by relation:  \mbox{$s_{12}+s_{23} + s_{31} = M^{2} +   m_{\alpha}^{2} + m_{\beta}^{2} + m_{\gamma}^{2}$} ($M$ and $m$'s label parent and final state particle masses respectively).   The total spin of two-particle subsystem is  labeled by $L$.     The cosine of    polar angle of  particle-1 in the rest frame of ($1(\alpha)2(\beta)$) system, \mbox{$z_{\alpha \beta} = \cos \theta_{\alpha \beta}$},   is given by ,
\begin{equation}
 z_{\alpha \beta}= - \frac{ s_{12}  (s_{23} - s_{31})  + (m_{\alpha}^{2} - m_{\beta}^{2}) ( M^{2} -m^{2}_{\gamma} ) }{2 M p_{\gamma}(s_{12}) 2 \sqrt{s_{12}} q_{\alpha \beta}(s_{12})},   \label{angz}
\end{equation}
where the momentum   factors $q$ and $p$ are defined by
\begin{align} 
 &  q_{\alpha \beta} (s_{12}) = \frac{\sqrt{ \left [ s_{12} -(m_{\alpha} - m_{ \beta})^{2} \right ]\left [ s_{12} -(m_{\alpha} + m_{ \beta})^{2} \right ] }}{2 \sqrt{s_{12}}}, \nonumber  \\
&  p_{\gamma} (s_{12})= \frac{\sqrt{ \left  [ s_{12} - (M -m_{\gamma})^{2} \right ] \left  [ s_{12} - (M +m_{\gamma})^{2} \right ]}}{2 M}. \label{pq}
\end{align}
Similarly, the other $z$'s are given by cyclically permutating sub- and super-indices  of Eqs.(\ref{angz}) and (\ref{pq}).
The dynamics of decay process    are described by   scalar functions $F$'s,  which only depend on subenergy of isobar pair ($s_{ij}$) by assumption.

 \begin{figure}
\includegraphics[width=0.48\textwidth]{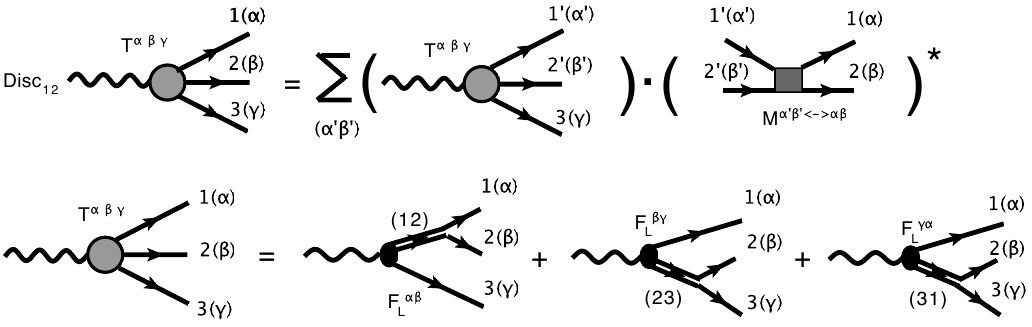}
\caption{    A diagrammatic representation of discontinuity relations in Eq.(\ref{discon}), the partial wave projection of Eq.(\ref{discon}) gives  Eq.(\ref{fdis}) . 
\label{isobarplot}}
\end{figure}

Considering the analytic properties of decay amplitude $T^{\alpha \beta \gamma}  $, the discontinuity crossing  unitarity   cut in  subenergy, {\it e.g.} $s_{12} $,  then reads,
\begin{align}\label{discon}
  \mbox{Disc}_{12}  T^{\alpha \beta \gamma} & (p_{1} , p_{2} ,p_{3}) \nonumber \\
 = & \frac{1}{2} \sum_{  (\alpha' \beta')   }     \int \frac{d^{3} \mathbf{ p}'_{1} }{(2\pi)^{3} 2 p^{'0}_{1}} \frac{d^{3} \mathbf{ p}'_{2} }{(2\pi)^{3} 2 p^{'0}_{2}}  \nonumber \\
  & \times   (2\pi)^{4 } \delta^{4} (p_{1}+p_{2} -p'_{1}-p'_{2})  \nonumber \\
   & \times    \mathcal{M}_{  \alpha \beta  \leftrightarrow \alpha' \beta' }^{*} (p_{1}p_{2} ; p'_{1} p'_{2})T^{\alpha' \beta' \gamma}  (p'_{1}, p'_{2} ,p_{3}) \nonumber \\
   +&  \sum_{L} (2 L+1) P_{L}(z_{\alpha \beta}) \sigma^{(\alpha \beta)}_{L}(s_{12}) ,
\end{align}
where the summation of $(\alpha' \beta')$ run over all allowed two-body intermediate states for $(1'(\alpha')2'(\beta'))$ pair,  last term $\sigma^{(\alpha \beta)}_{L}$ denotes the contribution from the rest of inelastic channels. In current work,  our discussion will be only limited to the three-body  subspace of inelastic channels by choosing \mbox{$\sigma^{(\alpha \beta)}_{L}=0$}. The partial wave expansion of two-body scattering amplitude $\mathcal{M}$ in a coupled-channel system  is given by
\begin{align}\label{scattamp}
& \mathcal{M}_{  \alpha \beta \leftrightarrow \alpha' \beta' } (p_{1}p_{2} ; p'_{1} p'_{2}) \nonumber \\
  &=   (16 \pi) \sum_{L}    (2 L+1)P_{L} (\cos \theta)  \left [t_{L} (s_{12}) \right ]_{(\alpha \beta) , (\alpha' \beta')} , 
\end{align}
 where $\theta$ is the angle between incoming and outgoing particles  of two-particle system.   The matrix $t_{L}$  denotes    coupled-channel  partial wave scattering amplitudes and it is normalized to \mbox{$\mbox{Im} t_{L}^{-1} = - \rho$}, where the  non-vanishing  elements of diagonal matrix $\rho$ are given by \mbox{$ \rho _{\alpha\beta } =  2 q_{\alpha \beta} /\sqrt{s_{12}}$}.  A diagrammatic representation of discontinuity relations in  Eq.(\ref{discon}) is shown in Fig.~\ref{isobarplot}.  Commonly, the kinematical  singularities are pulled out from decay  amplitudes by defining \mbox{$\widehat{F}^{(\alpha \beta)}_{L}(s_{12}) = F^{(\alpha \beta)}_{L}(s_{12})/\left [ \sqrt{s_{12}} q_{\alpha \beta} (s_{12}) p_{\gamma} (s_{12}) \right ]^{L} $} \cite{Guo:2010gx,Guo:2011aa}, where $\widehat{F}^{(\alpha \beta)}_{L}$ possess only dynamical unitarity cuts by assumption. The discontinuity relations for   scalar functions $\widehat{F}^{(\alpha \beta)}_{L}(s_{12})$  are  then derived from Eqs.(\ref{discon}) and (\ref{scattamp}),
\begin{align} 
\mbox{Disc}_{12}&  \widehat{F}^{(\alpha \beta)}_{ L}(s_{12})  \nonumber \\
=&  \sum_{ (\alpha' \beta' ) }  \frac{ \left [ t^{* }_{L}(s_{12})  \theta(s_{12} -s_{R}) \rho (s_{12})\right ]_{ (\alpha \beta) ,  (\alpha' \beta')}}{\left [ \sqrt{s_{12}} q_{\alpha \beta} (s_{12}) p_{\gamma} (s_{12}) \right ]^{L} }  \nonumber \\
   &\times \frac{1}{2}  \int_{-1}^{1} d z_{\alpha' \beta'}   P_{L}(z_{\alpha' \beta'} ) T^{ \alpha' \beta' \gamma} (s_{12}, s_{23},s_{31}),  \label{fdis}
\end{align}
where the non-vanishing   elements of diagonal matrix $s_{R}$  are \mbox{$s^{(\alpha \beta)}_{R} = (m_{\alpha } + m_{\beta})^{2}$}.  The self-consistent integral equation for $\widehat{F}^{(\alpha \beta)}_{L}(s_{12})$ is constructed by dispersion relation, 
\begin{equation}
 \widehat{F}^{(\alpha \beta)}_{L}(s_{12}) = \frac{1}{\pi} \int_{s^{(\alpha \beta)}_{R}}^{\infty} d s'  \frac{\mbox{Disc}_{12} \widehat{F}^{(\alpha \beta)}_{L}(s')}{s'-s_{12}}, \label{dispF}
 \end{equation}
where we have assumed that \mbox{$\mbox{Disc}_{12} \widehat{F}^{(\alpha \beta)}_{L}(\infty)=0$}, so that no subtractions is needed. The angular projection in Eq.(\ref{fdis}) has to be analytically continued when  discontinuity relation of $\widehat{F}_{L}$'s  is plugged into  Eq.(\ref{dispF}),  especially in the situation when  the dispersion integral runs out of physical decay region,   $z$'s are no longer  defined on real axis between $-1$ and $1$. The procedure of analytic continuation has been given in  \cite{Khuri:1960zz,Bronzan:1963kt,Aitchison:1965kt,Aitchison:1965dt,Aitchison:1966kt,Pasquier:1968zz,Pasquier:1969dt}. Similarly,   sets of equations for $\widehat{F}^{(  \beta \gamma)}_{L}(s_{23})$ and $\widehat{F}^{(\gamma \alpha  )}_{L}(s_{31})$ can be constructed  in   exactly the same approach, and together with Eq.(\ref{dispF}), they form a set of close coupled equations. The solutions of coupled-equation for $F$'s describe the three-body rescattering contribution    from both elastic and inelastic three-body channels. Eqs.(\ref{fdis}) and (\ref{dispF}) yield a basic representation of coupled-channel  formalism for three-body final state interaction. The rescattering effect is produced by exchanging particle between isobar pairs, and  the input of three-body equations are  the two-body scattering amplitudes, which may be obtained from experimental measurements.

  {\it Other representations of coupled-channel three-body formalism.}---Instead of solving Eqs.(\ref{fdis}) and (\ref{dispF}), we may also consider other representations of three-body equations, which demonstrate a explicit separation between rescattering contribution inside a pair and rescattering  between pairs.

As suggested in  \cite{Guo:2014vya}, first of all,  we may parametrize   amplitudes $\widehat{F}^{(\alpha \beta)}_{ L}$ by
  \begin{equation} 
   \widehat{F}^{(\alpha \beta)}_{L}(s_{12}) =   \left [\widehat{t}_{L} (s_{12}) g_{ L}  (s_{12}) \right ]_{(\alpha \beta) }, \label{fpara1}  
\end{equation}
where 
  \mbox{$ \left [ \widehat{t}_{L}   \right ]_{(\alpha \beta)  ,( \alpha' \beta')} = \left [ t_{L}  \right ]_{(\alpha \beta)  ,( \alpha' \beta')} /\left [q_{\alpha \beta} q_{ \alpha' \beta'}   \right ]^{L}$}. In general, $\widehat{t}_{L}$ has    both left-hand and right-hand   singularities, {\it i.e.}  \mbox{$  \mbox{Disc}  \widehat{t}_{L} =\theta(s - s_{L})    Im \widehat{t}_{L} + q^{2L}    \widehat{t}_{L}^{*  } \theta(s - s_{R})   \rho   \widehat{t}_{L}    $}, where $s_{L} $ labels   branch points of left-hand singularities. Therefore,  besides the unitarity cut,   the vector  $g_{L}$    has also   left-hand   singularities in order to keep $\widehat{F}$'s free off left-hand singularities, and  discontinuity relations for $g_{L}  $  thus  read,
 \begin{align} 
 & \mbox{Disc}_{12}\left[   g_{L} (s_{12})  \right ]_{(\alpha \beta)} \nonumber \\
 &= -          \left [   \widehat{t}_{L}^{* -1}(s_{12})   \theta(s_{L} - s_{12} )  \mbox{Im} \widehat{t}_{L}  (s_{12}) g_{L} (s_{12}) \right ]_{(\alpha \beta)}    \nonumber \\
  &     +       \frac{ q^{L}_{\alpha \beta} (s_{12}) }{  \left [ \sqrt{s_{12}}  p_{\gamma} (s_{12})  \right ]^{L} }   \theta(s_{12} -s^{( \alpha \beta)}_{R})   \rho_{\alpha \beta} (s_{12})  \nonumber \\
  & \quad \times \sum_{L'}   \frac{2L'+1}{2}     \int_{-1}^{1} d z_{\alpha \beta}P_{L}(z_{\alpha \beta} )  \nonumber \\
  & \quad    \times       \left [      P_{L'} (z_{\beta \gamma} )  F^{(\beta \gamma)}_{ L'} (s_{23})    +     P_{L'} (z_{\gamma \alpha} )  F^{(\gamma \alpha)}_{ L'} (s_{31}) \right ]  .  \label{gdis}
\end{align}
As   illustrated in single channel case in \cite{Guo:2014vya}, when the  discontinuity relation for $g$   is inserted into dispersion relation, the integral equations in two variables are obtained: one variable  is related to the angular projection; the other is associated  to  the dispersion integration. Fortunately,  the  Pasquier inversion technique \cite{Pasquier:1968zz,Aitchison:1978pw,Guo:2014vya} enable one to interchange the order of dispersive and angular integrations, and  eventually write a single     integral   equations for  $g_{L}$'s,
 \begin{align} 
 &  \left [ g_{L} (s_{12}) \right ]_{(\alpha \beta)}  \nonumber \\
 &= -        \frac{1}{\pi}   \int_{-\infty}^{ s^{(\alpha \beta)}_{L}}    \frac{d s'_{12}  }{s'_{12}-s_{12}}   \left [   \widehat{t}_{L}^{* -1}    \mbox{Im} \widehat{t}_{L}     g_{L} (s'_{12}) \right ]_{(\alpha \beta) }       \nonumber \\
  &  +       \sum_{L'}   \left [   \int_{ -\infty}^{(M-m_{\alpha})^{2} }  \!\!\!\!\!\!\!\!\!\!\!\!\!\!\!\!\!  ds_{ 23}   \ \ \mathcal{K}^{ (\alpha \beta ) \gamma \leftarrow (\beta \gamma) \alpha}_{g; LL'}(s_{12}, s_{23})     \right.\nonumber \\
  & \quad \quad \quad \quad \quad \quad \quad \quad  \quad \quad     \times  \left [ \widehat{t}_{L'}  (s_{23}) g_{ L'}  (s_{23} ) \right ]_{ \beta \gamma } \nonumber \\
  & \quad   \quad   \    +    \int_{-\infty}^{(M-m_{\beta})^{2}} \!\!\!\!\!\!\!\!\!\!\!\!\!\!\!\!\!  ds_{ 31}   \ \    \mathcal{K}^{(\alpha \beta) \gamma \leftarrow  (\gamma \alpha)\beta}_{g; LL'}(s_{12}, s_{31})  \nonumber \\
    & \quad \quad \quad \quad \quad \quad  \quad \quad   \quad \quad     \left.  \times  \left [ \widehat{t}_{L'}  (s_{31}) g_{ L'}  (s_{31}) \right ]_{ \gamma \alpha}  \right ]   ,  \label{pasqg}
\end{align}
where the kernel functions $ \mathcal{K}_{g}$'s are defined in  Eqs.(\ref{kernel1}- \ref{kernel2}), and $ \mathcal{K}_{g}$'s  do not  depend on  any dynamics but  only  on   kinematic factors.  Therefore,  the   'universality' properties of  $ \mathcal{K}_{g}$'s  allow one to compute them analytically,   which is a great advantage for numerical evaluation of Eq.(\ref{pasqg}). Similar equations for $g_{L}(s_{23})$ and $g_{L}(s_{31})$   are obtained by cyclic  permutation of both sub- and super-indices in Eq.(\ref{pasqg}).

Next, we consider another representation of three-body equations by parameterization of
\begin{equation} 
   \widehat{F}^{(\alpha \beta)}_{L}  (s_{12}) =    \left [\widehat{D}_{L}^{-1} (s_{12}) G_{ L} (s_{12})  \right ]_{(\alpha \beta)} , \label{fpara2}  
\end{equation}
where  $\widehat{D}_{L}^{-1}= \widehat{t}_{L} \widehat{N}_{L}^{-1}$ is denominator matrix functions of scattering amplitudes and has only right-hand   singularities by definition, and the left-hand singularities of $\widehat{t}_{L}$ are given by $ \widehat{N}_{L}$ matrix.  $\widehat{D}_{L}^{-1}$ and $\widehat{N}_{L}$  are simply a coupled-channel generalization of standard N/D method \cite{Chew:1960nd,Frye:1963nd}. In the single channel case,   function $\widehat{D}_{L}^{-1}$ may be referred to as the Muskhelishvili-Omn\'es (MO)  function \cite{Muskhelishvili,Omne}.
Thus, the vector $G$   possess only right-hand singularities and the discontinuity relations for $G_{ L}  $ read,
\begin{align} 
 & \mbox{Disc}_{12} \left [   G_{L} (s_{12})  \right ]_{(\alpha \beta)} \nonumber \\
 &=  \sum_{(\alpha' \beta')   }      \frac{ q^{L}_{\alpha' \beta'} (s_{12}) }{  \left [ \sqrt{s_{12}} p_{\gamma}(s_{12}) \right ]^{L}  } \nonumber \\
 & \quad \times  \left [ \widehat{N}_{L}^{*  } (s_{12})\theta(s_{12} -s_{R}) \rho(s_{12})  \right ]_{(\alpha \beta ), (\alpha' \beta')}  \nonumber \\
 &\quad \times  \sum_{L'}  \frac{2L'+1}{2}  \int_{-1}^{1} d z_{\alpha' \beta'} P_{L}(z_{\alpha' \beta'} )  \nonumber \\
  & \quad   \times  \left [       P_{L'} (z_{\beta' \gamma} )  F^{(\beta' \gamma)}_{ L'} (s_{23})   +     P_{L'} (z_{\gamma \alpha'} )  F^{(\gamma \alpha')}_{ L'} (s_{31}) \right ]  .  \label{Gdis}
\end{align}
Again, with the help of Pasquier inversion technique  \cite{Pasquier:1968zz,Aitchison:1978pw,Guo:2014vya}, a single integral equations for     $G_{ L} $ matrix are obtained,
\begin{align} 
 &   \left [ G_{L} (s_{12})  \right ]_{(\alpha \beta)} \nonumber \\
 &=     \sum_{(\alpha' \beta')   }  \sum_{L'}   \left [   \int_{ -\infty}^{(M-m_{\alpha'})^{2} }  \!\!\!\!\!\!\!\!\!\!\!\!\!\!\!\!\!   ds_{ 23}  \ \  \mathcal{K}^{ (\alpha \beta) \gamma \leftarrow \alpha'  (\beta' \gamma)}_{G;LL'}(s_{12}, s_{23})     \right.\nonumber \\
  & \quad \quad \quad \quad \quad \quad \quad \quad \quad    \quad      \times  \left [ \widehat{D}_{L'}^{ -1} (s_{23}) G_{ L'} (s_{23})   \right ]_{ ( \beta' \gamma)   }\nonumber \\
  & \quad   \quad  \quad   \quad   \ \     +    \int_{-\infty}^{(M-m_{\beta'})^{2}}  \!\!\!\!\!\!\!\!\!\!\!\!\!\!\!\!\!  ds_{ 31}  \ \   \mathcal{K}^{ (\alpha \beta) \gamma \leftarrow ( \gamma \alpha') \beta'}_{G;LL'}(s_{12}, s_{31})  \nonumber \\
    & \quad \quad \quad \quad \quad \quad  \quad    \quad \quad \quad          \left.  \times \left [ \widehat{D}_{L'}^{-1} (s_{31}) G_{ L'}  (s_{31}) \right ]_{ (\gamma \alpha')   }   \right ]   .  \label{pasqG}  
\end{align}
where  the kernel functions $ \mathcal{K}_{G}$'s, together with kernel function $ \mathcal{K}_{g}$'s defined in Eq.(\ref{pasqg}), are   given by
 \begin{align} 
 &  \mathcal{K}^{(\alpha \beta) \gamma \leftarrow \alpha' (\beta' \gamma)}_{g,G;LL'}(s_{12}, s_{23})   \nonumber \\
 &=   \frac{1}{\pi}  \left [ \theta(s_{23}) \int_{ s_{ \beta' \gamma}^{-}(s_{23}) }^{ s_{  \beta' \gamma}^{+}(s_{23}) } \!\!\!\!\!\!\!\!\!\!\!\!\!\!\! (C') \quad   - \theta(-s_{23})\int_{ s_{  \beta' \gamma}^{+}(s_{23}) }^{  \infty } \!\!\!\!\!\!\!\!\!\!\!\!\!\!\! (C') \quad \right ]    \frac{ d s'_{12} }{s'_{12}-s_{12}}   \nonumber \\
  &   \times        \frac{ q^{L}_{\alpha' \beta'} (s'_{12})   \left [ \sqrt{s_{23}} q_{\beta' \gamma} (s_{23}) p_{\alpha'}(s_{23}) \right ]^{L'} }{  M  p_{\gamma} (s'_{12})   \left [ \sqrt{s'_{12}}  p_{\gamma} (s'_{12})  \right ]^{L} }  \nonumber \\
  & \times    \left [ K_{L}^{(g,G) } (s'_{12}) \right ]_{(\alpha \beta ), (\alpha' \beta')}   \frac{2L'+1}{2}   P_{L}(z'_{\alpha' \beta'} )   P_{L'} (z'_{\beta' \gamma} )   ,  \label{kernel1} \\
  & \mathcal{K}^{(\alpha \beta) \gamma \leftarrow  ( \gamma \alpha') \beta'}_{ g,G; LL'}(s_{12}, s_{31})   \nonumber \\
 &=   \frac{1}{\pi}   \left [ \theta(s_{31}) \int_{ s_{\gamma \alpha'  }^{-}(s_{31}) }^{ s_{ \gamma \alpha'  }^{+}(s_{31}) } \!\!\!\!\!\!\!\!\!\!\!\!\!\!\! (C') \quad   - \theta(-s_{31})\int_{ s_{\gamma \alpha'  }^{+}(s_{31}) }^{  \infty } \!\!\!\!\!\!\!\!\!\!\!\!\!\!\! (C') \quad \right ]   \frac{ d s'_{12} }{s'_{12}-s_{12}}   \nonumber \\
   &   \times    \frac{ q^{L}_{\alpha' \beta'} (s'_{12})   \left [ \sqrt{s_{31}}  q_{ \gamma \alpha'} (s_{31}) p_{\beta'}(s_{31}) \right ]^{L'} }{  M p_{\gamma}  (s'_{12})  \left [ \sqrt{s'_{12}} p_{\gamma} (s'_{12}) \right ]^{L}  }    \nonumber \\
  & \times    \left [ K_{L}^{(g,G)  } (s'_{12}) \right ]_{(\alpha \beta ), (\alpha' \beta')}    \frac{2L'+1}{2}   P_{L}(z'_{\alpha' \beta'} )   P_{L'} (z'_{ \gamma \alpha'} )   , \label{kernel2}
\end{align}
where   the matrix $K^{(g,G)}_{ L}$ are given by \mbox{$  K^{(g)}_{ L}  (s'_{12}) = \mathbb{I} $} and  \mbox{$    K^{(G)}_{ L}  (s'_{12}) =   \widehat{N}_{L}^{*  } (s'_{12})  $} corresponding to $g_{L}$ and $G_{L}$   respectively.  The contour $C'$ is defined in Fig.~11 in \cite{Guo:2014vya}, and the integration limits (the boundary of Dalitz plot), {\it e.g.} \mbox{$ s^{\pm}_{  \beta \gamma} (s_{23}) $}, are given by
\begin{align}
 s^{\pm}_{  \beta \gamma} &(s_{23}) = \frac{M^{2} +   m_{\alpha}^{2} + m_{\beta}^{2} + m_{\gamma}^{2}   -s_{23}}{2}  \nonumber \\
& + \frac{(m_{\beta}^{2} - m_{\gamma}^{2}) (M^{2} - m_{\alpha}^{2})}{2 s_{23}} \pm   \frac{ 2 M p_{\alpha} (s_{23})   q_{\beta \gamma}(s_{23})}{ \sqrt{ s_{23}}}. \label{dalizbd}
\end{align}
Similar expression for $s^{\pm}_{\gamma\alpha}(s_{31})$ are obtained by cyclically permutating indices in Eq.(\ref{dalizbd}). As we see in Eqs.(\ref{kernel1}) and (\ref{kernel2}), the kernel functions $ \mathcal{K}_{G}$'s for $G_{L}$ equations  not only depend on dynamical functions $\widehat{N}_{L}$'s, but also has   off-diagonal contributions from rescattering between elastic and inelastic channels due to non-diagonal matrix $\widehat{N}_{L}$.  As for $g_{L}$ equations, although, the kernel functions $ \mathcal{K}_{g}$'s are totally diagonal, the off-diagonal contributions  appear in  the integral term over left hand cut (first term on the right-hand side of Eq.(\ref{pasqg})).  Unlike 'universal'  kernel functions $ \mathcal{K}_{g}$, because of  $\widehat{N}_{L}$ dependence in  kernel functions $ \mathcal{K}_{G}$, $ \mathcal{K}_{G}$'s now can only be computed  by numerical intergration in complex plane. 

Finally, the integral equations for $\widehat{F}$, $g$ and $G$ provide   three   equivalent representations of coupled-channel three-body formalism. As discussed in single channel three-body case in \cite{Guo:2014vya},   three different representations   in principle  yield the same result if $t_{L}$ matrix is well-defined   in complex plane.  In practice, the information of $t_{L}$ are usually only available in physical region on real axis, thus,   different approximate methods for solving dispersion integral equations are   used.   Therefore,     the difference in solutions from different representation      are expected  depending on the   approximations. In the single channel case \cite{Guo:2014vya}, different approximate methods by restricting the integration ranges seem    only change the overall normalization  of solutions in physical region and barely alter the resonance properties, so   the approximate solutions may be still justified. However, whether the conclusion still holds in coupled-channel case remains an open question.    Nevertheless,  single-integral-equation representations for $g$ and $G$   are clearly easier to solve numerically and more suitable for event by event based data analysis.

 {\it Summary.}---In summary, we derived sets of integral equations for coupled-channel three-body final state interactions based on the dispersion approach,   the formalism is presented in three different representations in Eqs.(\ref{dispF}), (\ref{pasqg}) and (\ref{pasqG}). 
 
 {\it Acknowledgement.}---We thank A.~P.~Szczepaniak for many fruitful discussions. 
  This research was supported in part   by the U.S.\ Department of Energy under Grant No.~DE-FG0287ER40365,  the Indiana University Collaborative Research Grant and  U.S.\ National Science Foundation under grant PHY-1205019.  We also   acknowledge support from U.S. Department of Energy contract DE-AC05-06OR23177, under which Jefferson Science Associates, LLC, manages and operates Jefferson Laboratory.

\appendix


\end{document}